\documentclass[aps,superscriptaddress,superbib,onecolumn,groupedaddress]{revtex4}

\usepackage[highlight]{neel}

\bibpunct{[}{]}{,}{n}{}{}

\graphicspath{{fig/}}

\usepackage[latin1]{inputenc}
\usepackage[T1]{fontenc}

\def\figurename{Fig.}

\begin{document}

\renewcommand\topfraction{0.8}
\renewcommand\bottomfraction{0.7}
\renewcommand\floatpagefraction{0.7}

\title{Head-to-head domain walls in one-dimensional nanostructures: an extended phase diagram ranging from strips to cylindrical wires}%

\author{S. Jamet}
\affiliation{Univ. Grenoble Alpes, Inst NEEL, F-38042 Grenoble, France}%
\affiliation{CNRS, Inst NEEL, F-38042 Grenoble, France}

\author{N. Rougemaille}
\affiliation{Univ. Grenoble Alpes, Inst NEEL, F-38042 Grenoble, France}%
\affiliation{CNRS, Inst NEEL, F-38042 Grenoble, France}

\author{J. C. Toussaint}
\affiliation{Univ. Grenoble Alpes, Inst NEEL, F-38042 Grenoble, France}%
\affiliation{CNRS, Inst NEEL, F-38042 Grenoble, France}

\author{O. Fruchart}
\affiliation{Univ. Grenoble Alpes, Inst NEEL, F-38042 Grenoble, France}%
\affiliation{CNRS, Inst NEEL, F-38042 Grenoble, France} \email[]{Olivier.Fruchart@neel.cnrs.fr}

\date{\today}


\begin{abstract}

So far magnetic domain walls in one-dimensional structures have been described theoretically only in the cases of flat strips, or cylindrical structures with a compact cross-section, either square or disk. Here we describe an extended phase diagram unifying the two pictures, extensively covering the (width,thickness) space. It is derived on the basis of symmetry and phase-transition arguments, and micromagnetic simulations. A simple classification of all domain walls in two varieties is proposed on the basis of their topology: either with a combined transverse/vortex character, or of the Bloch-point type. The exact arrangement of magnetization within each variety results mostly from the need to decrease dipolar energy, giving rise to asymmetric and curling structures. Numerical evaluators are introduced to quantify curling, and scaling laws are derived analytically for some of the iso-energy lines of the phase diagram.

\end{abstract}

\maketitle
\tableofcontents
\newpage


\newcommand{\Dtv}{$D_\mathrm{tw}$\/\xspace}
\newcommand{\Dbp}{$D_\mathrm{bp}$\/\xspace}
\newcommand{\Datw}{$D_\mathrm{atw}$\/\xspace}

\section{Introduction}

The investigation of magnetic domain walls~(DWs) in one-dimensional systems is an important topic in nanomagnetism, which has been active over about twenty years. Such DWs are associated with new physics dealing with their magnetization configuration at rest\cite{bib-MIC1997,bib-NAK2005}, their pinning with roughness and modulations of cross-section\cite{bib-CAY2004,bib-IVA2011,bib-SAL2013b}, their magnetization dynamics driven by an external field\cite{bib-BEA2005,bib-THI2006}, in-plane spin-polarized currents\cite{bib-GRO2003b,bib-THO2006,bib-THI2008} or  perpendicular spin currents of either tunnel or spin-Hall origin\cite{bib-KHV2009,bib-SEO2012,bib-KHV2013,bib-EMO2013}. They have also been proposed as a basis for low-power devices with logic or memory functionalities, based on DW motion\cite{bib-ALL2005,bib-PAR2008}. Low-dimensional structures also provide an opportunity to use surface-related effects such as perpendicular anisotropy and the Dzyaloshinskii-Moriya interaction etc. Here we will however consider structures with all dimensions larger than a few nanometers, so these effects may be disregarded. We will focus on low-anisotropy, soft magnetic materials, \ie displaying so called \hth or \ttt DWs.

One may consider two different types of one-dimensional systems. Using a top-down approach, thin films and lithography can be combined, producing mostly strips, \ie with a rectangular and rather flat cross-section. Strips are ideal for physics and devices thanks to their versatility of fabrication and easiness of inspection. DWs in strips have been investigated theoretically and experimentally extensively over the past twenty years. In a bottom-up approach, pores are formed in, \eg, irradiated polycarbonate or anodized alumina, which can be filled by electroplating\cite{bib-FER1999a}. This yields wires, by which we mean with a compact cross-section such as disk or square. More complex structures such as tubes and core-shell can also be produced by atomic layer deposition\cite{bib-DAU2007,bib-CHO2010} or electroplating\cite{bib-WAN2008c,bib-PRO2012}. There exists a recent review covering both the variety of synthesis strategies and magnetic properties of such magnetic structures. The consideration of DWs in such wires started later than for strips, typically about ten years ago, and so far mostly models and simulations are available\cite{bib-SOU2014}. These first considered wires and later on tubes (with an empty or non-magnetic core). Experimental results are emerging only now\cite{bib-BIZ2013,bib-FER2013,bib-FRU2014}, however a rising interest is expected as wires is the natural geometry for possibly realizing a three-dimensional race-track memory\cite{bib-PAR2008}.

Strips and wires have mostly been considered so far as different systems. Thus there has not been a global thinking about the names of DWs in these. This sometimes yields conflicting statements or naming such as the multiple names Bloch-point wall~(BPW)\cite{bib-THI2006,bib-KIM2013}, vortex wall~(VW)\cite{bib-FOR2002,bib-WIE2004a,bib-HER2004a} or pseudo-vortex wall\cite{bib-OTA2013} for the same object in a wire. Also, the latter should not be confused with the vortex wall in strips\cite{bib-MIC1997,bib-NAK2005}, which has a different shape and even topology, as we will detail later. This conflicting name of a so-called vortex wall for walls with distinct topologies in a wire versus in a strip is an issue as a wire with, e.g., a square cross-section, may be obtained continuously by increasing the thickness and/or decreasing the width of a strip. Making this connexion is not a purely theoretical consideration, as processes are emerging for producing wires with a well-defined rectangular or square cross-section\cite{bib-SER2014}. Finally, due to the complexity of displaying and understanding three-dimensional configurations such as found in wires and thick strips, the description of these is sometimes not free of misconceptions, so that a simple description and classification would be useful to guide their understanding. For these two reasons it is desirable to sketch a global phase diagram of DWs in 1D systems. This is the purpose of the present work.

In the following we first propose in \secref{sec-overview} a short overview of the current understanding of domain walls in one-dimensional systems. In \secref{sec-sketching} we then sketch the extended phase diagram based on general considerations and symmetry arguments, and in \secref{sec-simulations} we refine it based on micromagnetic simulations. The simulations also allow us to discuss the key aspects of DWs in this picture, so at to deliver a simple view and means of classification in terms of a small number of features. In \secref{sec-scaling-laws} a few analytical scaling laws are derived to highlight the physics at play in the phase diagram. These resulting sections have been written to be readable mostly as stand-alone sections, depending on the interest of the reader.

\section{A short overview of existing knowledge}
\label{sec-overview}

As said above, we consider low-anisotropy, soft magnetic materials and disregard surface effects. In micromagnetics these materials may be described by the sole values of their magnetization $\Ms$ and exchange stiffness~$A$. With these we define the dipolar exchange length $\DipolarExchangeLength=\sqrt{2A/\muZero\Ms^2}$, which can be used to scale the present results to any soft magnetic material~(for $\mathrm{Fe}_{20}\mathrm{Ni}_{80}$, so-called permalloy, $\DipolarExchangeLength\approx\lengthnm{5}$). Besides, in absence of effects such as exchange bias the time-reversal symmetry applies, to that the physics of \hth and \ttt DWs are identical. In the following we will designate these walls by \hth, for the sake of concision.

Before addressing one-dimensional structures, let us recall the basics of DWs in three- and two-dimensional systems, \ie in bulk and thin-film materials, still in rather soft magnetic materials. This is interesting as some phenomena arising in these, have a similarity with those that we will describe in one-dimensional systems. Besides, identical words are used in both cases such as \textsl{vortex wall} or \textsl{asymmetry}. They are used to describe phenomena sharing common features, however with a different geometry. It is important to review their use in 3d and 2d systems to avoid any confusion with the 1d case.

When a sample is subdivided in several domains, be it for the sake of decreasing magnetostatic energy or because of magnetic history, DWs are present at the domain boundaries. As a general rule DWs arrange themselves to avoid the formation of magnetic charges, which would otherwise give rise to magnetostatic energy and thus to a cost of dipolar energy. In the case of $\angledeg{180}$ DWs this results in so-called Bloch DWs, in which magnetization lies in the plane of the wall. More generally, the plane of a domain wall tends to bisect the direction of magnetization in the neighboring two domains. The component of magnetization perpendicular to the wall remains uniform from one domain to the other through the wall, while the field of perpendicular component is similar to a $\angledeg{180}$ Bloch wall. This way the volume density of magnetic charges $-\mathrm{div}\vectM$ is zero. The picture is more complex in thin films, where magnetic charges may arise at surfaces with an areal density $\vectM \dotproduct \vect n$ with $\vect n$ the outward normal to the surface. For very thin films perpendicular magnetization cannot be sustained as in a Bloch wall, and magnetization rotates in the plane over a length scale equal to the domain wall width. This is the geometry of the N\'{e}el wall\cite{bib-NEE1955}. In thicker films domain walls remain of Bloch type deep inside the film, while at surfaces magnetization tends to turn parallel to the surface to avoid the formation of surface charges\cite{bib-HUB1969,bib-LAB1969}. These features were called later N\'{e}el caps\cite{bib-FOS1996}, due to the profile of surface magnetization being very similar to that of N\'{e}el walls. A refinement of this picture is the spontaneous shift of N\'{e}el caps along the direction perpendicular to the domain wall plane, so as asymptotically avoid any volume charge even in the vicinity of N\'{e}el caps. This arrangement was called an \textsl{asymmetric} Bloch wall\cite{bib-LAB1969}. When the thickness of the film is reduced the N\'{e}el caps at opposite surfaces become close one to another, so that the cross-section of the domain wall looks like a swirl of magnetization. This feature, taking place to close as much as possible the magnetic flux, has been named a \textsl{vortex wall}\cite{bib-HUB1998b}. It should not be confused with the so-called vortex wall occurring in thin strips. The history of domain and domain wall structure resulting from the principle of pole avoidance was recently reviewed by A.~S.~Arrott\cite{bib-ARR2011}.

\begin{figure}
  \begin{center}
  \includegraphics[width=159.595mm]{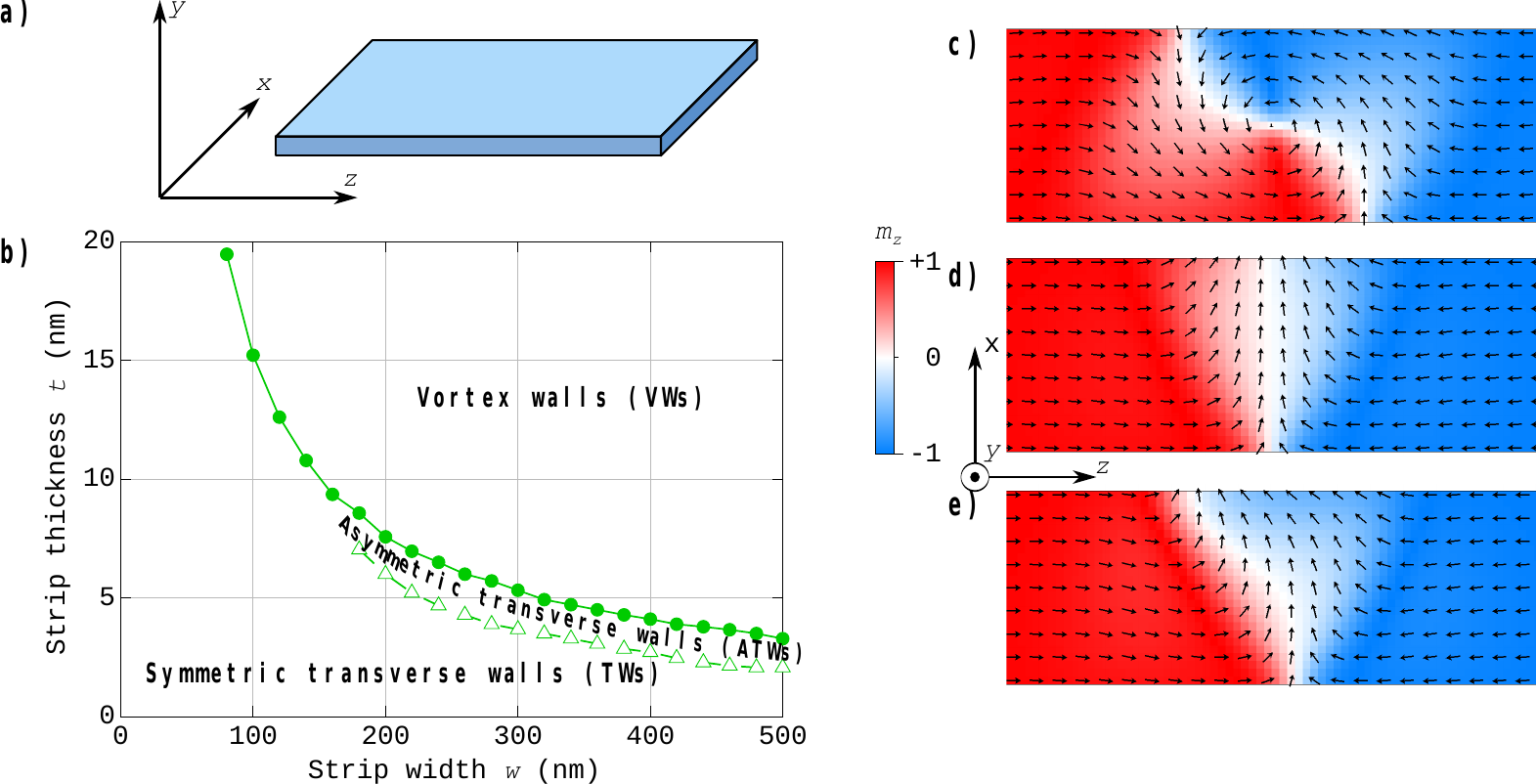}%
  \caption{\label{fig-tw-vw-strip}\textbf{Domain walls in strips} (a)~Coordinates used to describe strips: $x$ and $y$ along transverse directions, and $z$ along the strip length (b)~Phase diagram of DWs in strips known to-date\cite{bib-MIC1997,bib-NAK2005}. Labels indicate the state of lowest energy among TW, ATW and VW\cite{bib-NAK2005}. (c-e)~Mid-height views of micromagnetic simulations of the vortex~(VW), transverse~(TW) and asymmetric transverse~(ATW) walls. The strip width is $\unit[100]{\nano\meter}$ and thickness $\unit[16]{\nano\meter}$ for c and d, and $\unit[28]{\nano\meter}$ for e. The color codes the magnetization component along~$z$.}
  \end{center}
\end{figure}

Domain walls in strips are known to take two forms: the transverse wall~(TW) and the VW, as initially described by McMichael and M. Donahue\cite{bib-MIC1997}. In both cases magnetization remains mostly in-plane. In the TW magnetization inside the wall is directed along the in-plane direction transverse to the strip\bracketsubfigref{fig-tw-vw-strip}{d}. Its shape is roughly triangular, reminiscent of the formation of $\angledeg{90}$ walls in thin films to avoid having locally net charges, associated with a long-range magnetostatic energy. In a VW the magnetization is curling within the plane around a small area with perpendicular magnetization called the vortex core, of diameter $\approx3\DipolarExchangeLength$\bracketsubfigref{fig-tw-vw-strip}{c}\cite{bib-FEL1965b}. Examination of the detailed magnetic microstructure reveals again $\angledeg{90}$ sub-walls. Both types of DWs have the same total charge $2tw\Ms$, with $t$~and $w$ the strip thickness and width, respectively. Simulations and magnetic force microscopy provide the picture of a rather uniform distribution of these charges over the entire area of the wall\cite{bib-CHA2010}, an argument which we will use in the scaling laws section. TWs and VWs have been investigated numerically up to $w=\lengthnm{500}$ and $t=\lengthnm{20}$. The TW is of lower energy than the VW for $tw\lesssim61\DipolarExchangeLength^2$, and of higher energy for larger width or thickness\bracketsubfigref{fig-tw-vw-strip}{b}\cite{bib-NAK2005}. Both DWs persist as a metastable state in a large part of the ground-state domain of the other DW. In experiments it is often the metastable TW which is observed in the region of stable VW, following its initial stability during preparation under a transverse applied field\cite{bib-KLAe2004,bib-CHA2010}. A refinement revealed later is that the TW undergoes a transition towards large and wide strips, from a symmetric shape to an asymmetric one\bracketsubfigref{fig-tw-vw-strip}{e}. The resulting wall was named an asymmetric transverse wall, which we will write ATW\cite{bib-NAK2005}.

\begin{figure}
  \begin{center}
  \includegraphics[width=160.056mm]{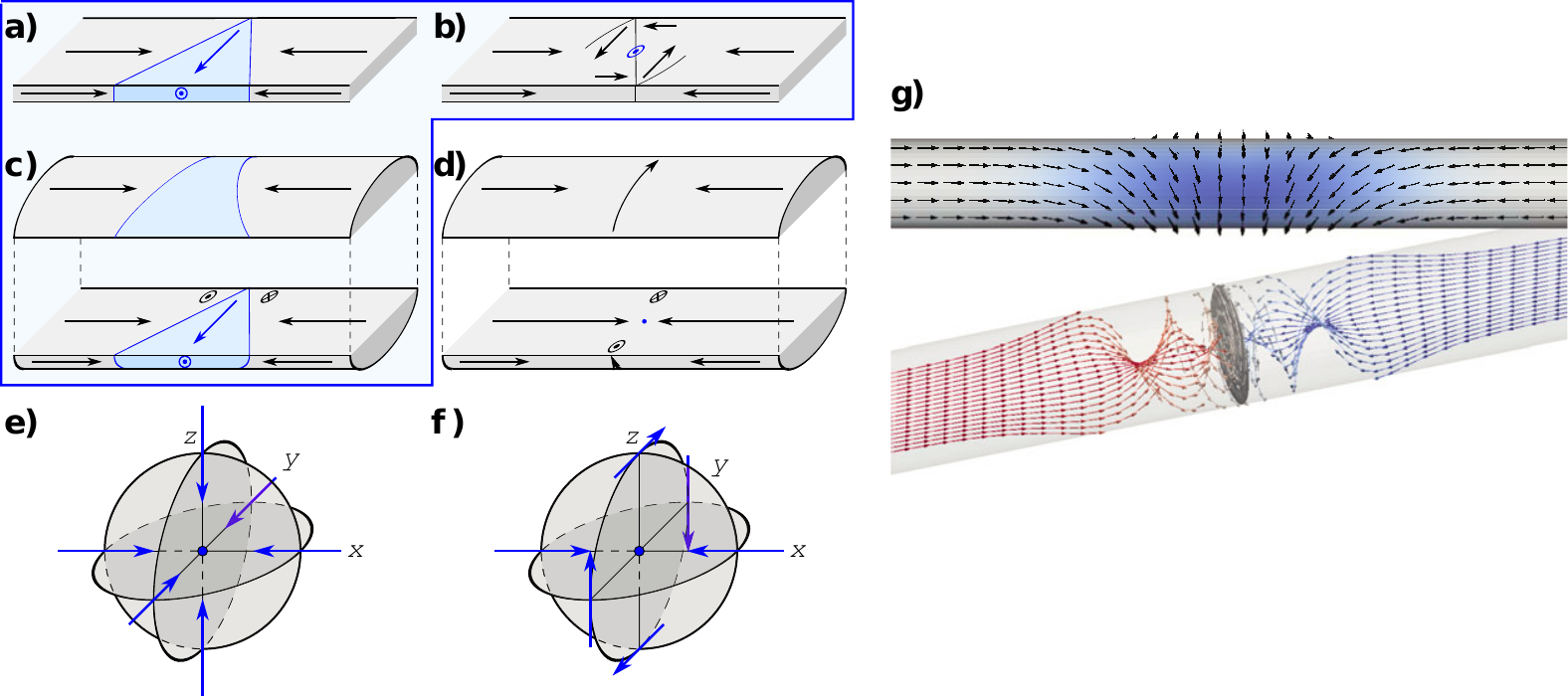}%
  \caption{\label{fig-bpw-wire}\textbf{Bloch points and domain walls in one-dimensional structures} (a)~Transverse wall~(TW) and (b)~Vortex wall~(VW) in a strip. (c)~TW and (d)~Bloch-point wall~(BPW) in a cylindrical wire (e-f)~Bloch-points, of types hedgehog and curling. (g)~Micromagnetic simulations of a BPW. The top view shows surface magnetization. The bottom one highlights streamlines of magnetization, while the central disk is the surface where $m_z=0$.}
  \end{center}
\end{figure}

Letting aside early numerical approaches\cite{bib-ARR1979,bib-ARR1979b}, DWs have been investigated thoroughly in cylindrical structures only later, pioneered by R.~Hertel\cite{bib-HER2002a,bib-NIE2002} and H.~Forster\cite{bib-FOR2002,bib-FOR2002b}. Two types of DWs were predicted. The first type of DW was named TW in analogy with the case of strips, because its core displayed a significant component of magnetization along a direction transverse to the wire\bracketsubfigref{fig-bpw-wire}{a;c}. It is again the most stable form for low lateral dimension, either diameter for disk section or side for square section, typically below $\approx7\DipolarExchangeLength$\cite{bib-FOR2002,bib-WIE2004a,bib-THI2006}. The main feature of the second type of DW is again a curling of magnetization as for a VW in a strip. However in the wire case the curling occurs around the axis, allowing a three-dimensional flux-closure\bracketsubfigref{fig-bpw-wire}{b;d}. This curling is made possible in a wire, compared to a strip, because exchange energy is not prohibitively costly along the thickness due to the somewhat large dimension. This orthoradial curling makes it impossible to sustain a radial component of magnetization on the wire axis. Besides, as the \hth nature of the DW forbids sustaining a longitudinal component either in the core of the DW, there must exist a point where the magnitude of the magnetization vector vanishes. This quite unusual object is called a Bloch point, predicted and described in the early days of micromagnetism\bracketsubfigref{fig-bpw-wire}{e;f}\cite{bib-FEL1965,bib-DOE1968,bib-MAL1979,bib-KLE1983}. Its reality is accepted although it has not been imaged directly due to its very small size; only observing extended boundary conditions allows to infer its existence and monitor its dynamics\cite{bib-HAR1973,bib-MAL1979,bib-KAB1989}. Due to the curling this domain wall was initially named a VW, a name still in wide use today. The name Bloch-point wall~(BPW) was proposed later by A.~Thiaville and Y.~Nakatani\cite{bib-THI2006} to avoid the confusion with the strip case. Some other authors call it a pseudo-VW, stressing that this name is intended to avoid confusion with a VW in a strip\cite{bib-OTA2013}. In this manuscript we will stick to the name BPW for this reason. Finally, DWs were later studied numerically and analytically in tubes\cite{bib-USO2007,bib-ESC2007,bib-LAN2007}. The physics is very similar to the case of wires, with the essential difference of the removal of a core of magnetization along the axis, so that the equivalent of a BPW consists mostly of an orthoradial curling, with no Bloch point. It was also mentioned that TWs undergo a distortion upon increasing the wire diameter. This effect has been named \textsl{helical domain wall}\cite{bib-CH2012} or \textsl{pseudo transverse wall}\cite{bib-BIZ2013} depending on the authors.

In the above we introduced the terms circulation and curling. These are interchangeable, the latter bearing the meaning of the curl operator and related to circulation of magnetization along a closed path (Stokes theorem). The name curling has been used for a long time in magnetism to describe structures with such circulation, introduced in nucleation theory as what is now known as the curling model\cite{bib-FRE1957}. Notice that curling is not necessarily a transient feature during magnetization switching: curling structures are known to be relevant for magnetization textures at rest\cite{bib-HUB1999}. The notion of curling is more general than that of vortex, which often has a meaning related to a particular distribution of magnetization such as for the VW in a flat strip. Also, the name vortex gives more importance to its core structure, such as in fluid swirls. In the course of this manuscript we will encounter weakly curling structures, for which there exists no well-defined core. Going along this line, we could use vortex state to describe the flux-closure in a flat disk, however curling would be more suited for a ring. Similarly, in a nanotube we could say longitudinal curling wall for the equivalent of Bloch-point wall for a wire, as the core is removed.

\section{Sketching the phase diagram}
\label{sec-sketching}

\subsection{Preliminary discussion}

It is not our idea to deliver a comprehensive view of all details of domain walls in one-dimensional structures. The first reason is that this would be a task formidable and difficult to summarize, due to the large variety of geometries to be considered, \eg, with cross-sections of type rectangular, disk and elliptical, hollow (for tubes) etc. The second reason is that many fine micromagnetic features appear as the size of a system is increased, such as edge and corner effects, configurational anisotropy\cite{bib-COW1998c,bib-COW1998b}; the task would be even more formidable. In fact, not all these features are crucial to understand the energetics and possibly the dynamics of the walls. Simplifying the picture by selecting the most important features is therefore desirable. For instance, it should be of use to understand and analyze outputs of micromagnetic simulations in simple words, whereas they are often difficult to display and discuss because of their possibly three-dimensional character.

Most of the concepts discussed and simulations reported here, consider one-dimensional structures with a rectangular cross-section. This choice stems from the whish to span continuously the panorama from strips to wires with an experimentally-relevant geometry. Nevertheless, figures and features are also discussed for a circular cross-section, which so far is the most relevant geometry for wires. As considering domain walls in tubes instead of wires was shown to change numbers however not the general texture of the walls\cite{bib-USO2007,bib-ESC2007,bib-LAN2007}, the case of tubes could probably be extrapolated from the present work, missing only the precise position of the various features in the diagram.

We will build a phase diagram consisting of lines defining regions where such or such domain wall is stable, metastable or does not exist. We will name \textsl{first-order} those lines separating two different states, each existing and being stable or metastable on either side of the line. For instance, the well-known iso-energy line between the TW and the VW in strips is of first order. On graphs these will be depicted with a bold line, full for separating the two states of lowest energy, dotted when the ground state is not one of the two states considered. We will name \textsl{second-order} those lines associated with the continuous rise of an internal degree of freedom in a DW (characterized by an ordering parameter), associated with a breaking of symmetry. In this case only one type of DW exists on either side, and there is no metastability. For instance, this is the case in strips for the transition of the TW to the ATW. Their features have been described in detail for magnetic systems\cite{bib-HUB1999}. On graphs these will be depicted with thin lines, full or dotted with the same meaning as indicated above. The features expected for a transition depending on its order are summarized in~\tabref{tab-transitions}.

Due to the symmetry upon exchanging $t$ and $w$, any line has its counterpart symmetric with respect to the diagonal $t=w$, be it of first or second order. In particular, if a line is itself not symmetric with respect to the diagonal, and therefore is not perpendicular to it at the intercept, this implies the existence of two distinct lines intercepting at the same point on the diagonal, related to configurations that can be obtained one from another through a rotation of $\pi/2$ around the axis. We will call \textsl{lower triangle} the part for $t\leq w$, and \textsl{upper triangle} its symmetric counterpart for $t\geq w$. $x$ and $y$ are coordinates along the directions transverse to the wire, while $z$ is used along the wire\bracketsubfigref{fig-tw-vw-strip}a.

Finally it has been said that we restrict the discussion to soft magnetic materials. The discussion and calculations are not material-dependant, once scaled with the dipolar exchange length~$\DipolarExchangeLength$. In the following we use general arguments to derive a simplified sketch for the phase diagram, such as symmetry, phase transition of first or second order, topology and scaling laws (the detail of the latter being found in \secref{sec-scaling-laws}). In \secref{sec-simulations} we use micromagnetic simulation to give a firmer basis to the sketch, and refine quantitatively its various elements. We also define and calculate estimators for a few quantities characterizing the internal degrees of freedom, and which may serve as ordering parameters when describing second-order transitions.

\begin{table}
  \centering
  \label{tab-transitions}
  \bgroup
  \def\arraystretch{1.5}
  \begin{tabular}{lm{0.5cm}m{5cm}m{0.5cm}m{5cm}}
    \hline\hline
     && First order && Second order \\\hline
    Stability && Two states on either side, stable or metastable && Only one state on either side of the transition. No metastability. \\
    Order parameter && Abrupt transition && Continuous rise; breaking of symmetry \\
    Energy && \includegraphics[width=22.780mm]{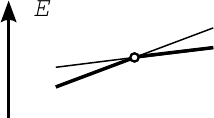} &&  \includegraphics[width=22.101mm]{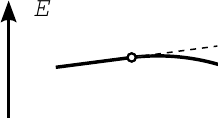} \\\hline\hline
  \end{tabular}
  \egroup
  \caption{Common features for first and second order transitions, which we will use to describe the transitions from one micromagnetic state to another}
\end{table}

\subsection{Transverse versus vortex walls}

\begin{figure}
  \begin{center}
  \includegraphics[width=150.104mm]{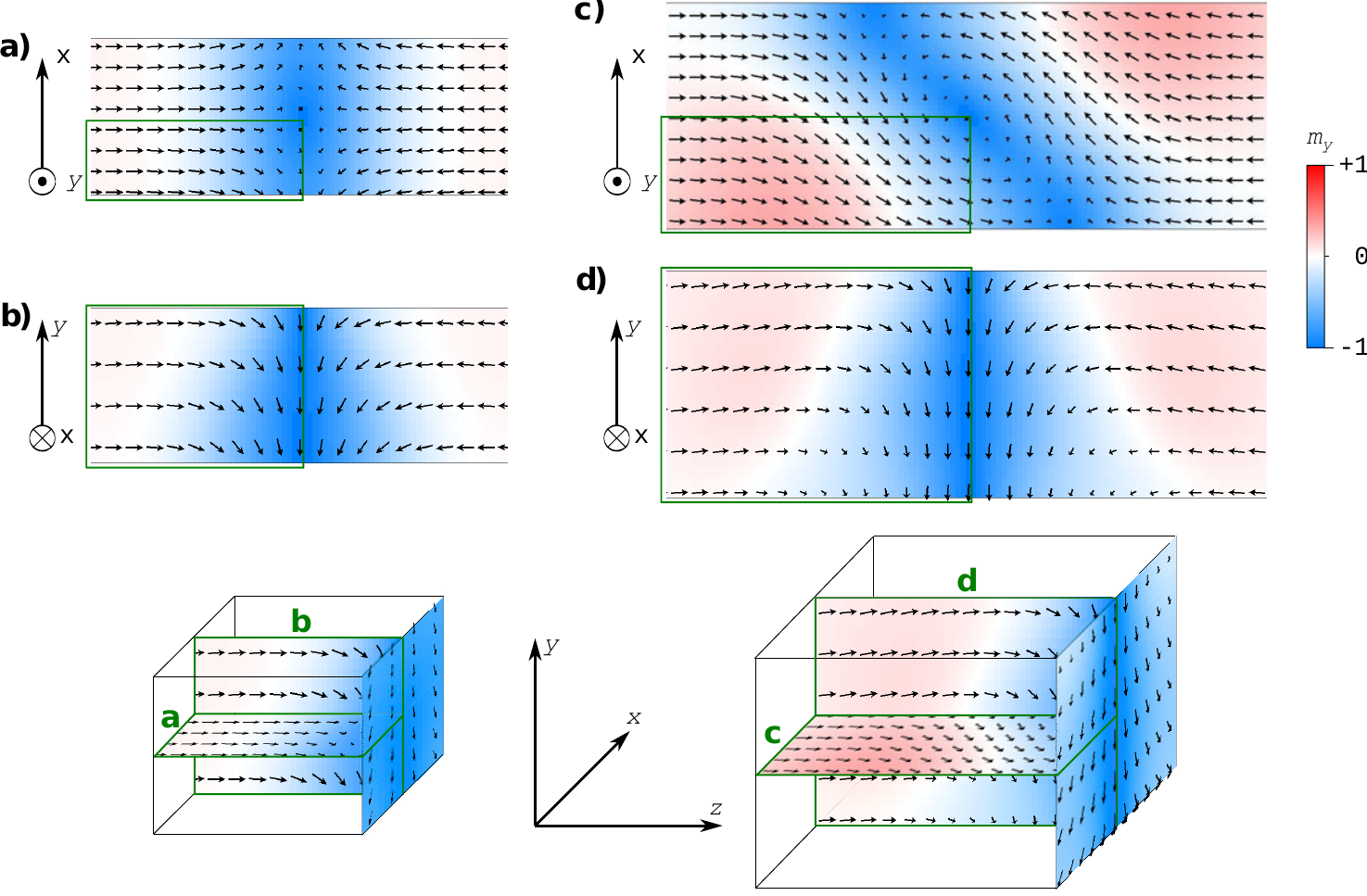}%
  \caption{\label{fig-tw-vw-square}\textbf{Rise of curling for transverse-vortex walls (TVWs) in wires with square cross-section}. Wires of side (a,b)~$\unit[30]{\nano\meter}$ and (c,d)~$\unit[44]{\nano\meter}$. (a;c)~show mid-height top views, while (b,d)~show mid-depth side views. In both cases the color codes magnetization along $y$, \ie the transverse component of the DW. At the bottom are shown open 3D views, for which the areas displayed are framed in the views above.}
  \end{center}
\end{figure}

Let us examine in more detail the structure of the TW and VW in strips. In both cases a tube of magnetization goes through the strip: from edge to edge along $x$ for the TW, from top to bottom along $y$ for the VW. Thus, as a view of mind it is possible through a continuous deformation of the magnetic texture to transform a TW into a VW, following a path in the phase space. In other words, these two DWs  share the same \textsl{topology}. Considering a TW and a VW for a given rectangular cross-section, let us imagine that we continuously change the geometry of the strip towards a square cross-section. When $t=w$ the two types of walls should be degenerate and be obtained one from another through a rotation of $\pi/2$ around the wire axis. This is illustrated on \figref{fig-tw-vw-square}, based on micromagnetic simulations. These show that a wall for a square cross-section indeed displays both a transverse feature~(a flux of magnetization from one side to the opposite one, \subfigref{fig-tw-vw-square}d) and a vortex feature~(circulation of magnetization around the transverse component, \subfigref{fig-tw-vw-square}c), in the common sense used in strips for a TW and a VW. This is analogous to a liquid-gas transition, which determines a line of first-order transition, however where a path may be found to go continuously from one to the other around a critical point~(here: through the critical line~$t=w$). We therefore propose to name this family of walls the transverse-vortex wall, or mixed transverse-vortex wall~(TVW). The similarity of both types of walls has already been outlined by some authors, stating \eg \textsl{the
vertical Bloch line is probably the one that represents closest similarity with a transverse domain wall}\cite{bib-HER2014}. Of course, when one feature is dominating the other and no confusion is possible, restricting the name to TW or VW is desirable for clarity. We may also add information when necessary to avoid confusion, about the direction of the flux of magnetization: a $x$-TW when the transverse component is along $x$ as in flat strips, or a $y$-TW (or $y$-TVW) when the transverse component is along $y$, as may be considered for small cross-section\bracketsubfigref{fig-tw-vw-square}{a-b}. In this case there is little variation of the direction of magnetization within a cross-section, which make it a near realization of the 1d model\cite{bib-THI2006}.

\begin{figure}
  \begin{center}
  \hspace*{-2mm}\includegraphics[width=159.965mm]{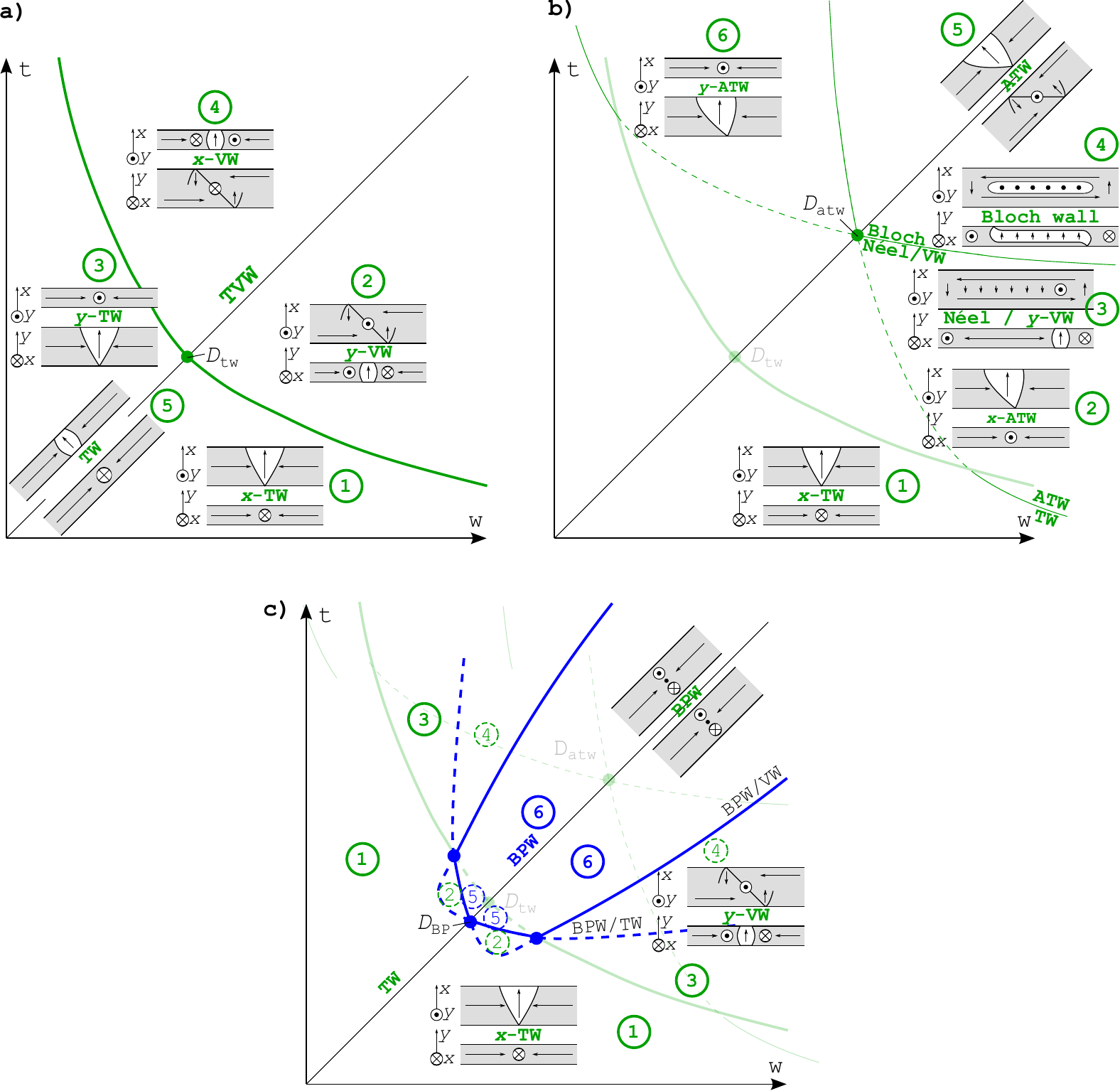}%
  \caption{\label{fig-diagrams}\textbf{Expected phase diagram}. The sketches are based on existing knowledge extended by symmetry arguments. Bold (resp. thin) lines mark first-(resp. second) order transitions. Full lines are used when the ground state is one of the two states considered, while dotted lines are used when both states are metastable, the ground state being of another variety. The phase diagram is built from a to c, adding one by one different types of transitions lines (a)~First-order line separating transverse walls~(TWs) and vortex walls~(VWs) (b)~Second-order lines separating symmetric/asymmetric TWs, and also VWs walls from Bloch-type walls (also called Landau-type, see text).}
  \end{center}
\end{figure}

Let us come back to the known first-order line separating TVWs in strip with transverse component along $x$ and $y$ respectively, \ie with predominantly transverse~(TW) and vortex~(VW) feature. This is depicted on \subfigref{fig-diagrams}a. In the lower triangle usually displayed, $x$-TW are the most stable in region~1, and $y$-VW are in region~2. Considering points symmetric when swapping $t$ with $w$, a mostly $x$-TW is transformed into a mostly $y$-TW, and it is straightforward that this line is symmetric with respect to the diagonal. There is one single TW/VW line in the diagram, and it crosses the diagonal at right angle at a point we will name \Dtv. In region~3 it is the $y$-TW which is of lowest energy, and in region~4 it is the $x$-VW. This may be checked rigorously with the following argument. Let us name $\Delta \mathcal{E}(t,w)$ the difference of energy between the $x$-TW and the $y$-VW at a given point in the phase space. $\Delta \mathcal{E}$ is of class $C^2$ for $(t,w)>0$ so that we may consider its Taylor expansion around~\Dtv. $\Delta \mathcal{E}$ remains zero along both the diagonal and the iso-energy line, while being non-zero along other directions pointing into stability/metastability regions, which from the expansion requires that both lines are perpendicular. Note that the symmetry with respect to the diagonal is fortunately consistent with the phenomenological law $wt\approx\mathrm{Cte}$ already reported\cite{bib-MIC1997,bib-NAK2005}~(see also the scaling law in \secref{sec-scaling-laws}).

\begin{figure}
  \begin{center}
  \hspace*{-2mm}\includegraphics[width=160mm]{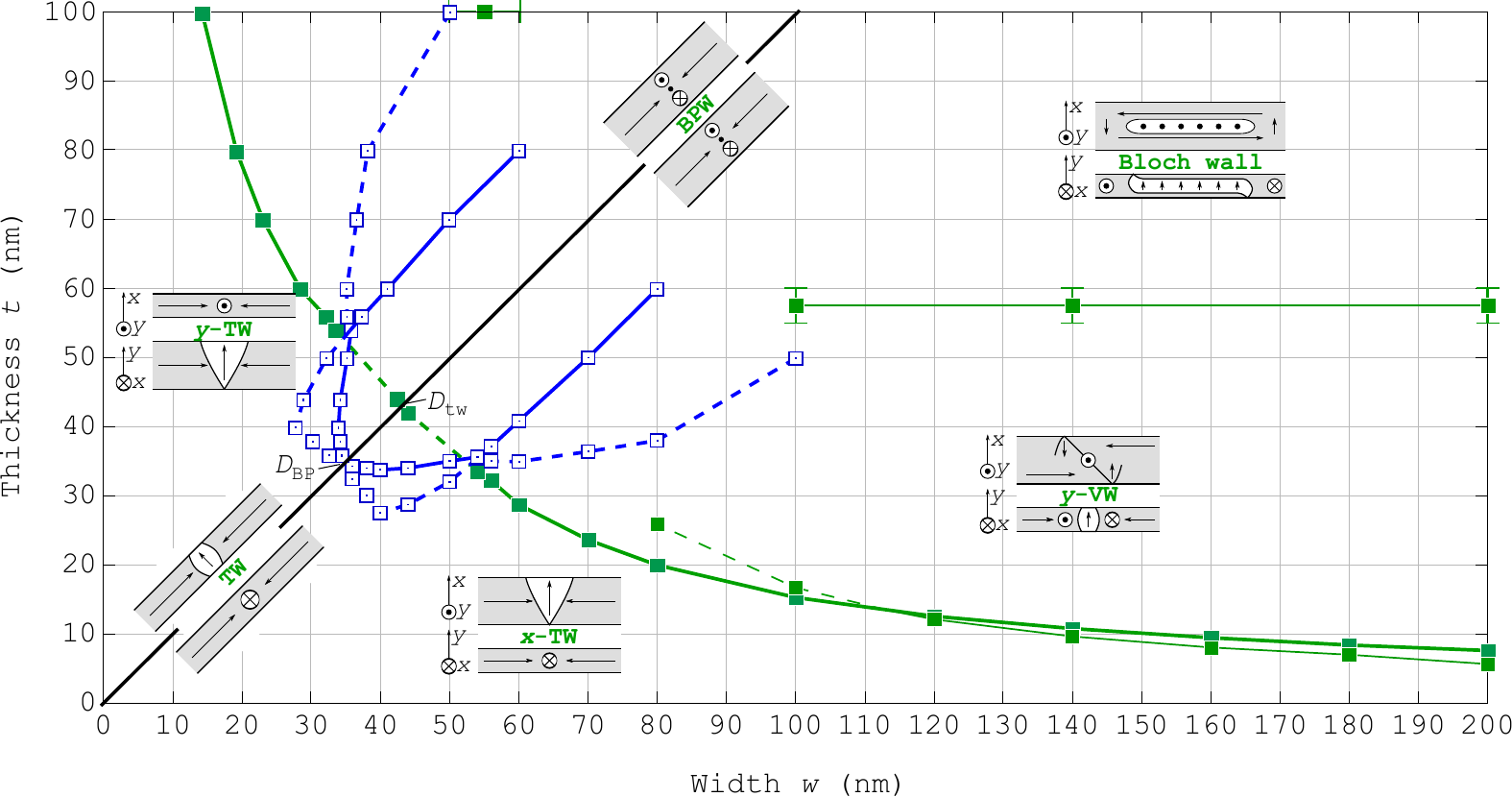}%
  \caption{\label{fig-diagram-simuls}\textbf{Phase diagram refined with micromagnetic simulations}. The notations for the various lines is the same as in \figref{fig-diagrams}.}
  \end{center}
\end{figure}

\subsection{Asymmetries}

The following discussion is illustrated on \subfigref{fig-diagrams}b. In the preliminary discussion we mentioned the second-order transition from a $x$-TW~(labeled~1) to an $x$-ATW~(labeled~2) upon increase of, \eg, the thickness of a strip. This asymmetry towards a slanted wall is reminiscent for the zigzag domain walls found in extended thin films with uniaxial anisotropy, as already pointed out in \cite{bib-NAK2005}. The asymmetry is characterized by the fact that the locus of entry and outlet of the flux of magnetization on either edge of the strip are shifted along the length of the strip\bracketsubfigref{fig-tw-vw-strip}e. Let us assume that this second-order line exists for a broad range of geometries, and we follow it towards the diagonal and beyond, in the upper left triangle. In this region the broad and narrow dimensions of the rectangular section are interchanged with respect to its symmetric lower-right triangular region, so that starting from a DW with an $x$-through-flux along the long transverse direction and thus with a dominant transverse character, we end up with a DW still with a $x$-through-flux however now along the short transverse direction and thus with a dominant $x$-VW character. The tube of flux is then entering from one flat surface, which may be called bottom, and exiting from the other flat surface, say top. Swapped back into the lower triangle, this is a $y$-VW wall, labeled~3. Introducing an asymmetry between these two entry/outlet locus transforms the vortex into a Bloch wall of finite length terminated by a surface vortex at either of its ends, labeled~4. This feature of topology is well known for finite-length Bloch walls\cite{bib-ARR1997,bib-HER1999,bib-FRU2009b,bib-FRU2010c}. This occurrence of asymmetry is thus related to the physics of the transition from a N\'{e}el wall to a Bloch wall upon, \eg, applying a transverse magnetic field\cite{bib-MID1963,bib-HUB1998b,bib-FRU2009b}. In thin films, it has also been recognized to be of second order\cite{bib-RAM1996}.

As the diagonal $t=w$ plays \apriori no role for this line of second-order transition, there is no reason why it should be perpendicular to it. Its symmetric with respect to the diagonal should therefore be a distinct line, which shall be also added to the phase diagram. In this view of mind, on one side of the diagonal it pertains to the TW/ATW transition, while on the other side it pertains to the transition from N\'{e}el or vortex, to the Bloch wall. When wrapped, these two lines are found one above the other on each side, and intercepting the diagonal at the same point\bracketsubfigref{fig-diagrams}b.

\begin{figure}
  \begin{center}
  \hspace*{-1mm}
  \includegraphics[width=181.729mm]{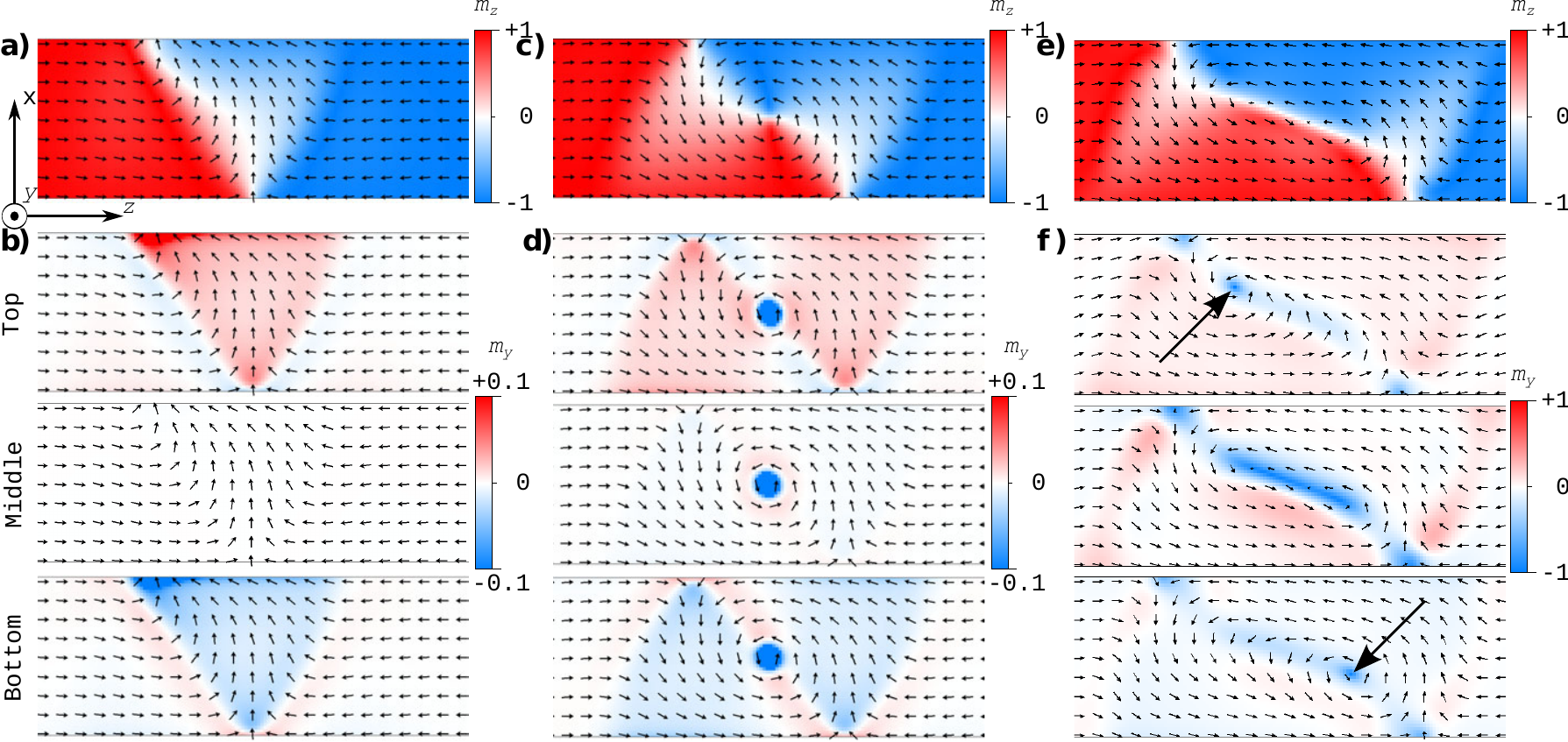}%
  \caption{\label{fig-landau-wall}\textbf{The various domain walls for thick strips}. Views of (a,b)~TW, (c,d)~VW and (e,f)~Landau walls in strips of width $\unit[200]{\nano\meter}$. The thickness is $\unit[16]{\nano\meter}$ from a to d, and {\unit[60]{\nano\meter}} for e and f. (a,c,e)~display mid-height views with color coding $m_z$. (b,d,f)~display top, middle and bottom views with color coding $m_x$. In b and d the color contrast is enhanced tenfold, so that $|m_x|=0.1$ uses full contrast. In f the large black arrows point at the surface vortices where the flux of magnetization from the Bloch wall flows from the strip.}
  \end{center}
\end{figure}

If we focus again on the lower triangle~($w>t$), the evolution with thickness of the magnetic microstructure of lowest energy is now the following: for low thickness we shall consider TWs~(label~1), then for increasing thickness: possibly ATWs~(label~2), VWs, and finally the transition from vortex to Bloch wall~(labels 3 to~4). This latest transition has been described in detail recently\cite{bib-FRU2015}. The occurrence of the Bloch wall allows to increase the length of the entire \hth wall by giving it an internal structure similar to a tilted Landau flux-closure pattern\bracketsubfigref{fig-landau-wall}{e,f}. This again is a clear illustration of the similarity of VW and TW. The Landau-type DW has been described above as an evolution of the VW\bracketsubfigref{fig-landau-wall}{c,d}. Similarly, it can be viewed as the evolution of an ATW\bracketsubfigref{fig-landau-wall}{a,b}, the in-plane transverse part having initially a N\'{e}el type, being converted to a Bloch type due to the large thickness. In this phase diagram, asymmetric walls are expected for $t=w$ beyond \Datw. We will see that micromagnetic simulations will refine this picture.

\subsection{Bloch-point walls}

The situation of BPWs versus TVWs is depicted on \subfigref{fig-diagrams}c. So far the BPW has been discussed in the literature only for the geometry of wires, \ie for $t=w$. It was predicted for a disk\cite{bib-HER2002a,bib-FOR2002} as well as a square\cite{bib-THI2006} cross-section. The experimental confirmation of it existence was only provided recently, based on a disk cross-section\cite{bib-FRU2014}. The TVW and the BPW have a different topology. They are separated by an energy barrier and are (meta)stable over a large range of diameters. Along the diagonal they are thus associated with a first-order transition at a point that we will call \Dbp\bracketsubfigref{fig-diagrams}c, previously determined to be located at $w=t\approx7\DipolarExchangeLength$. Moving along the diagonal from \Dbp towards larger diameters, the energy of the BPW becomes lower than that of the TVW\cite{bib-THI2006}. As the energy of domain walls is continuous as a function of strip dimensions, this implies that there should exist a region on either side of the diagonal where the BPW is the wall of lowest energy, associated with first-order transition lines that we outline in more details below.

Outside the diagonal a $x$-TVW~(towards a dominant TW feature in the lower triangle) and a $y$-TVW~(towards a dominant VW feature in the lower triangle) have a different energy, so that their must exist two distinct first-order lines: one for the $x$-TW-BPW transition~(could be called TW-BPW in the lower triangle), another one for the $y$-TVW-BPW transition~(could be called VW-BPW in the lower triangle). On the diagonal the TW and VW are degenerate, so both lines should intercept there, at \Dbp. As regards this intercept, noticing that upon crossing the diagonal a VW becomes a TW and vice versa, the two lines are mirror one to another with respect to the diagonal. As the energy difference between the TVW and the BPW are anywhere differentiable with class $C^2$, no kink is expected at $D$. Therefore, the two curves display complementary angles with the diagonal.

Appending these curves on the phase diagram concerning transitions in the family of TVW requires a further discussion. If \Dbp were found for a larger diameter than \Dtv, it is probable that none of these curves intercept the TW/VW first-order transition line. Consequently in the lower triangle it would always be the $y$-VW which is of lower energy under the TVW/BPW lines, so that the one relevant for determining the ground state would be the VW/BPW one\bracketfigref{fig-diagram-simuls}. Conversely if \Dbp is found for a smaller diameter than \Dtv, then the relevant line for determining the ground state is a portion of the TW/BPW one for small width, and the VW/BPW for larger width\bracketfigref{fig-diagram-simuls}. Micromagnetic simulations will show that this is the relevant situation. In that case, notice that the two iso-energy lines intercept at a triple point the TW/VW line: when a TW and a VW have the same energy, their energy difference with that of a BPW is the same, in particular when it is vanishing. To the contrary, there is no reason why \Dbp and \Dtv should be identical. On \subfigref{fig-diagrams}c the energy of the various domain walls is ordered the following way: TW<VW<BPW in region~1, TW<BPW<VW in region~2, VW<TW<BPW in region~3, VW<BPW<TW in region~4, BPW<TW<VW in region~5, and BPW<VW<TW in region~6.

\section{Micromagnetic simulations}
\label{sec-simulations}

\subsection{Methods}

Micromagnetic simulations have been performed with the parameters of Permalloy $\mathrm{Ni}_{80}\mathrm{Fe}_{20}$: $\muZero\Ms=\unit[1.0053]{\tesla}$ for the spontaneous magnetization, $A=\unit[10]{\pico\joule\per\meter}$ for the exchange stiffness, and zero magnetocrystalline anisotropy. The resulting dipolar exchange length is $\DipolarExchangeLength=\sqrt{2A/\muZero\Ms^2}\approx\unit[5]{\nano\meter}$. Although directly applicable to Py, our simulations are relevant to any other magnetic material with no magnetocrystalline anisotropy, scaling all lengths with $\DipolarExchangeLength$.

We used two micromagnetic codes. The first code is the finite differences OOMMF\cite{bib-OOMMF}. It has been used to cover the phase diagram continuously from the case of flat strips with a rectangular cross-section, to wires with a square cross-section. The cell size was $\unit[1\times 1\times 2]{\nano\meter\cubed}$ for strips of width smaller than $\unit[60]{\nano\meter}$, $\unit[2\times 2\times 2]{\nano\meter\cubed}$ for strips with higher widths up to $\unit[100]{\nano\meter}$, and $\unit[4\times 4\times 2]{\nano\meter\cubed}$ above. Damping was set to~1 to speed up convergence, with no consequence on the result as we are interested in equilibrium states. Magnetic moments at the extremities of the strips are fixed to avoid the formation of end domains, and the length of the strips is chosen such that the aspect ratio is at least~$10$. The second code is feellgood, a home-built code based on a finite element scheme, \ie using tetrahedra to discretize matter. It is thus better suited to describe curved boundaries, and was used to investigate wires. Both circular and square cross-sections were considered. The size of tetrahedra was about \unit[4]{\nano\meter}. The Landau-Lifshitz-Gilbert equations are integrated using an original semi-implicit scheme which is consistent up to order two in time and unconditionally stable. It combines a linear inner iteration with a renormalization stage for the nodal magnetization for which a rigorous proof of convergence was established\cite{bib-ALO2012,bib-ALO2014}. The computation of the magnetostatic interactions constitutes the major bottleneck
in the efficiency of micromagnetic codes. The solution adopted here uses a time-optimized and parallelized version of the Fast Multiple Method running on multi-core conventional processors\cite{bib-MES2012,bib-ScalFmm}. The computing time devoted to magnetostatic interactions then becomes comparable to the time required for the other steps in the calculation.

\subsection{First-order transitions}

Simulations were first used to test the existence and refine the location of the first-order equilibrium lines\bracketfigref{fig-diagram-simuls}. Series of simulations of the two states to be compared were performed for a given strip width~(along~$x$) and variable thickness~(along $y$), and the thickness-dependent energy was fitted with a second-order polynomial. For first-order transitions (TW/VW, TW/BPW and VW/BPW) states of either type are (meta)stable on both sides of the equilibrium line, and at the transition their energy curves cross each other with different slopes. Thus, an accurate determination of the crossing point is straightforward.

The VW/TW iso-energy line was already known for $w\geq\unit[80]{\nano\meter}$, although computed mostly in a 2D micromagnetic scheme so far\cite{bib-MIC1997,bib-NAK2005}. Our 3D calculations essentially confirm this data, and extend the line towards the diagonal. We confirm the intercept of the latter at right angle derived from symmetry arguments, and this for a square side $\approx\unit[43]{\nano\meter}$, or $\approx8.6\DipolarExchangeLength$. This shows that the phenomenological scaling law $tw\lesssim61\DipolarExchangeLength^2$, derived in the range of width $\unit[80\mathrm{-}500]{\nano\meter}$, remains of high accuracy up to the diagonal.

\begin{figure}
  \begin{center}
  \includegraphics[width=165.180mm]{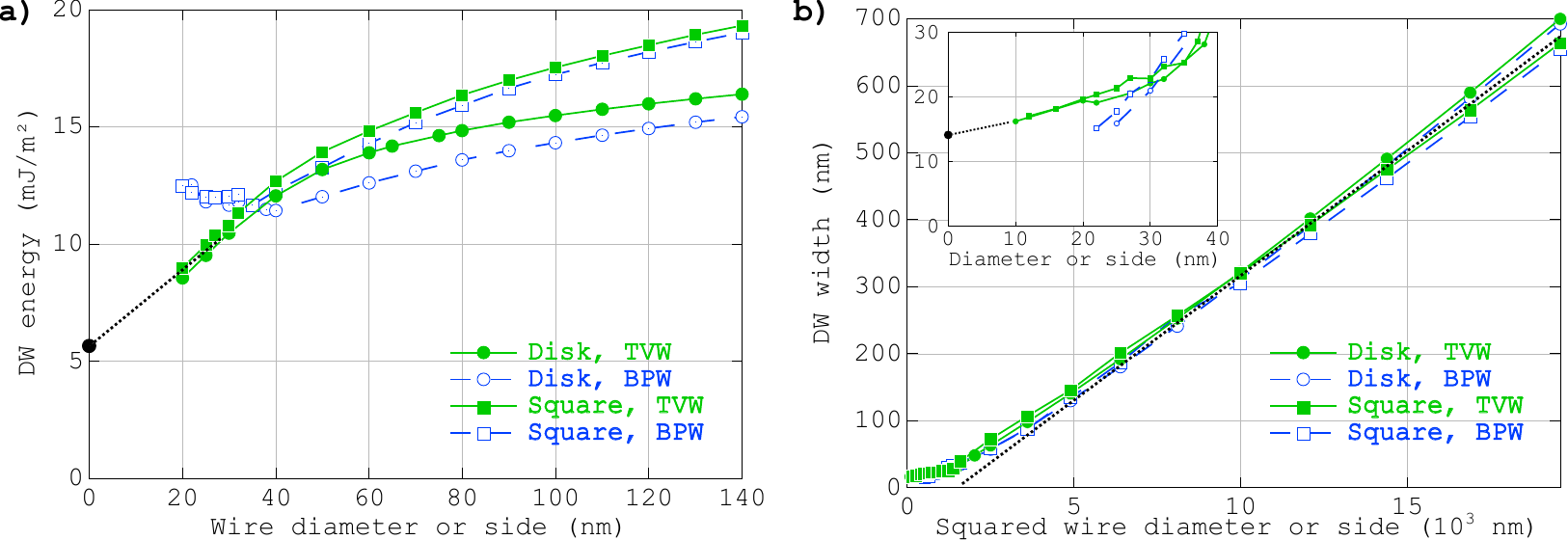}%
  \caption{\label{fig-energy-wires}\textbf{DW energy and width for square- and disk-based wires} (a)~DW energy normalized to cross-section. The dot on the $y$ axis stands for the limit $4\sqrt{A\Kd/2}\approx\unit[5.7]{\milli\joule\per\meter}$~\cite{bib-THI2006}. (b)~DW width versus squared (main graph) or lateral dimension (inset). The dot on the $y$ axis of inset stands for the limit $2\sqrt{2A/\Kd}\approx\unit[14.1]{\nano\meter}$~\cite{bib-THI2006}. Narrow-dotted lines are guides to the eye. Notice that for a given value $d$ the area of the cross-section is larger by the amount $4/\pi$ in squares, compared to disks.}
  \end{center}
\end{figure}

So far BPWs had only been considered for wires. The iso-energy points had been determined to be $\approx6.2\DipolarExchangeLength$ for disk\cite{bib-FOR2002} and $\approx7.0\DipolarExchangeLength$ for square\cite{bib-THI2006} cross-sections. We confirm the magnitude of $\approx7.0\DipolarExchangeLength$ for the transition in square cross-section, and slightly lower for the disk cross-section. \figref{fig-energy-wires} shows the energy of the two types of walls for both types of cross-sections, plotting the energy normalized to the section area. This normalization, following Ref.\cite{bib-THI2006}, has the advantage of yielding a quantity converging to a finite value towards low diameter. From this graphs it is clear that the stability of BPWs with respect to TVWs is enhanced in wires with a disk cross-sections, compared to those with a square cross section. The reason for this will be discussed in the next section. Besides this, our simulations confirm that for larger dimensions the BPW remains stable in a significant range of anisotropic cross-sections, over both walls with a dominant TW or VW feature, \ie cores along the long or short transverse dimension. For instance, for $w=\unit[100]{\nano\meter}$ the BPW remains of lower energy than the VW in the range $t\in\unit[[50\mathrm{-}100]]{\nano\meter}$. Finally, the existence of two distinct lines for TW/BPW and VW/BPW transitions is confirmed, with a triple point when meeting the VW/TW transition line.

\subsection{Second-order transitions and estimators for their parameter}

For first-order transitions the two states considered were separated by an energy barrier of finite height, and differed in orientation or even in topology. To the contrary, across a second-order transition a symmetric state continuously develops a feature breaking its symmetry. There are two possibilities to determine the transition in this case. The first one is based on energies. The difference between the energy of one state with the extrapolation of the other one (extrapolation because the two states do not exist at the same spot in phase space), is expected to scale with $(t-t_\mathrm{c})^{1/2}$ in the Landau theory with $m^2$ and $m^4$ terms; $t_\mathrm{c}$ is the thickness for transition and $m$ an order parameter characterizing the magnitude of the breaking of symmetry. This $1/2$ exponent was already reported for second -order transition in micromagnetics, despite the complexity of the system considered\cite{bib-HUB1999}. One may then fit with parabola both branches. These parabola should intercept and share the same slope at the transition, so that the latter cannot be determined easily. In practice, closely-spaced data points and a high accuracy in the numerics are required. Second, it is possible to monitor the order parameter $m$. Extrapolating it towards zero provides the locus of the transition. An exponent $1/2$ is also expected from the Landau theory with $m^2$ and $m^4$ terms. Such an exponent has already been reported\cite{bib-HUB1999,bib-FRU1999b,bib-FRU2001}, while other exponents such as $1/6$ seem to be possible\cite{bib-NAK2005}. We followed the procedure based on energy.

\begin{figure}
  \begin{center}
  \includegraphics[width=164.781mm]{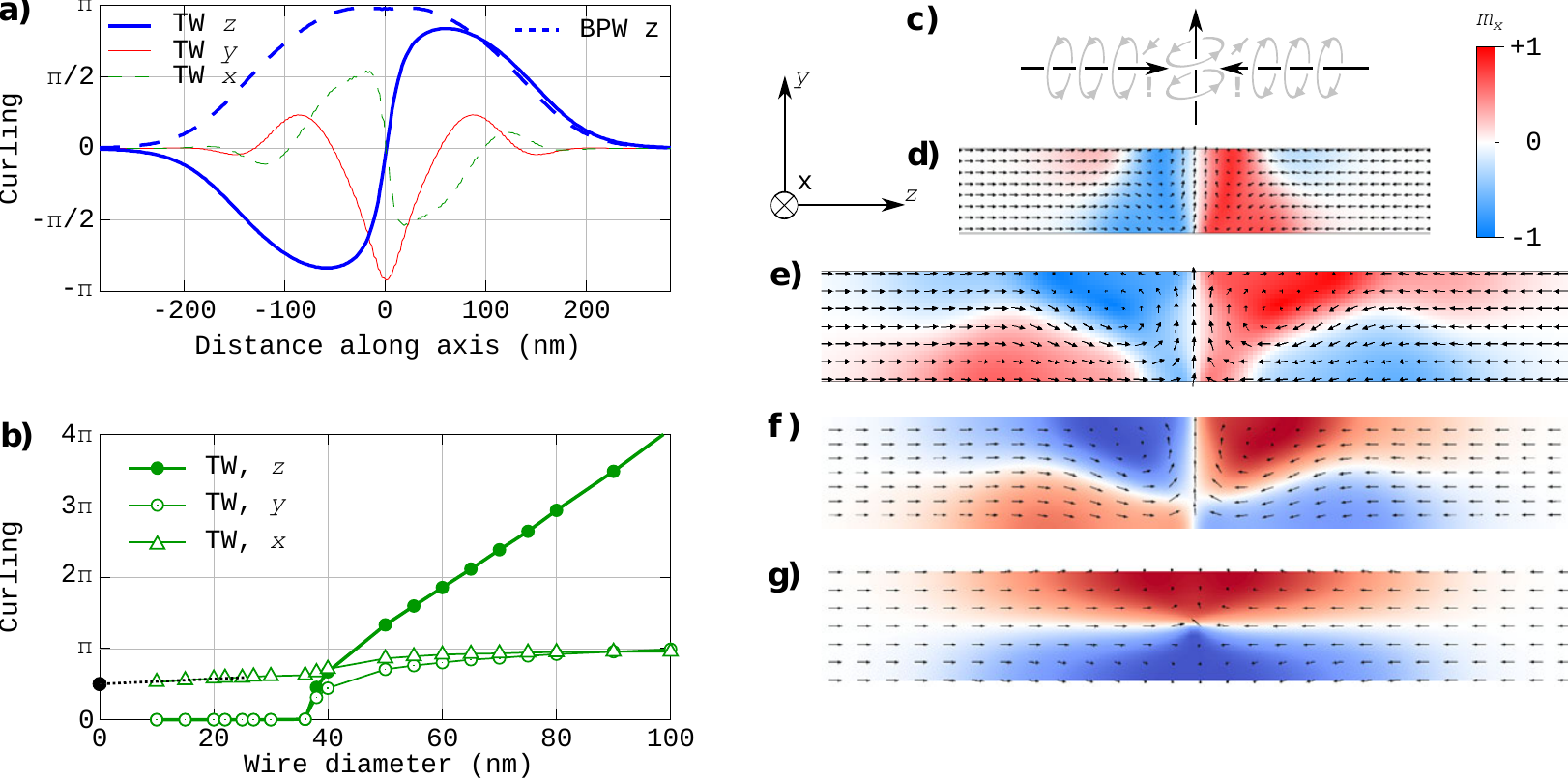}%
  \caption{\label{fig-curling}\textbf{Curling features}. (a)~the three components of the layer-resolved curling for a $y$-TVW, and longitudinal component for a BPW, in a wire with disk section of diameter $\unit[80]{\nano\meter}$ (b)~integration along the wire of the absolute value of layer-resolved curling for a $y$-TVW in a wire with disk section (c)~Illustration of the competition of transverse with longitudinal curling for TVWs. The dark arrows stand for the mean direction of magnetization, while the light arrows stand for the curling part. Frustrated areas are highlighted with an exclamation mark. Cross-sections highlighting curling through coloring the $x$~component for a $y$-TVW DW, for (d)~$\unit[60\times60]{\nano\meter}$ and (e)~$\unit[80\times80]{\nano\meter}$ square cross-sections, and (f)~\unit[80]{\nano\meter}-diameter disk cross-section of a $y$-TVW. The sign of transverse curling is the same in all cases. (g)~Longitudinal curling around a BPW for a \unit[80]{\nano\meter}-diameter disk cross-section.}
  \end{center}
\end{figure}

The first case of breaking of symmetry is a one-dimensional asymmetry, such as for the TW-ATW transition for rather thin strips. The examination of energy around the transition confirms that it is of second order. Previously, the transition line had been outlined for strip widths $\unit[180]{\nano\meter}$ and higher\cite{bib-NAK2005}. The present simulations extended it for smaller widths, and show that it rises faster than the TW-VW first-order line, and even cross it around $w=\unit[115]{\nano\meter}$. This will be explained with simple arguments with a scaling law\bracketsecref{sec-scaling-laws}.

Simulations confirm that a breaking of symmetry also occurs for thick strips, where the vortex at the center of the VW is transformed into a Bloch wall of finite length\bracketsubfigref{fig-landau-wall}{e,f}. For strip width $\unit[100]{\nano\meter}$ or higher this line is essentially flat, and located at $t=\unit[57.5\pm2.5]{\nano\meter}$. In \secref{sec-sketching} we argued that this line may be viewed as the symmetric of the prolongation of the TW/ATW, and that they may intercept on the diagonal. We searched along the diagonal when a TVW may become asymmetric, with entry and exit points of the magnetization flux at different longitudinal positions. Such configurations have not been found, neither for square nor disk cross-sections, and this up to side or diameter equal to $\unit[140]{nm}$. The reason why no TW asymmetry is present for wires may be because a more efficient way of reducing magnetostatic energy is developed before, as described below.

We now turn to a third type of second-order transition, which had been described previously in the context of near-single-domain particles\cite{bib-HUB1999}, however not in the context of domain walls. The well-known VWs in flat strips are characterized by the curling of magnetization around the $y$-transverse component, \ie the core of the vortex. In a system of very small size this curling cannot occur, because it would have a prohibitive cost in exchange. Accordingly, it is known and we confirm that for wires of small diameter the magnetization profile of a TW is essentially one-dimensional\bracketsubfigref{fig-tw-vw-square}{a,b}. For larger diameter curling may be achieved through the continuous orthoradial deformation of an initially symmetric volume, either clockwise or anti-clockwise, and as such may give rise to a second-order transition. Circulation may be proposed as an order parameter, for example any component of the quantity $\Curl\vect{m}$. When integrated on a disk cross-section and normalized with~$d$, $(\Curl\vect{m})\dotproduct{\vect u}_z$ should equal~$\pi$ for a perfectly orthoradial vector field, \ie close to the situation found at the center of a BPW. The three components of the z-resolved estimator for a TVW in a disk-based wire are plotted against its diameter in \bracketfigref{fig-curling}{a}, where the transverse component $y$ is chosen to be the azimuth of the core of the TVW: this is a $y$-TVW. Transverse curling adds a vortex character to the domain wall as seen from $(\Curl\vect{m})\dotproduct{\vect u}_y$. This DW has thus well-developed transverse and vortex features, visible depending on the cross-section examined~(\subfigref{fig-tw-vw-square}{c,d} and \subfigref{fig-curling}{d-f}).
When integrated in absolute value over the length of the wire to avoid cancellation~(see \subfigref{fig-curling}a), and normalized with $d^2$, a curling quantity again of the order of $\pi$ is expected for a wall of length~$d$. The exact value depends on both the strength of curling inside the wall, and the wall width. Plotting it versus wire diameter allows to investigate the transition\bracketsubfigref{fig-curling}b. Starting from zero for small diameter, the integrated circulation grows rapidly beyond  $d=\unit[36]{\nano\meter}$~($\approx7\DipolarExchangeLength$) consistent with a second-order transition, and this simultaneously along the longitudinal and $y$~transverse directions. The integrated $x$ curling is non-zero for small diameter and starting from roughly $\pi/2$, because the layer-resolved curling is not zero: each side of the wall having the shape of a quarter vortex whose core would be at the wire side~(see \subfigref{fig-tw-vw-square}{b,d}). Despite this non-zero background the superimposed second-order-like rise is clearly visible. Finally, while longitudinal curling keeps rising due to the increasing wall length, as discussed later, both transverse components tend to a value close to $\pi$, consistent with the picture of a vortex wall. Going back to strips, $\approx7\DipolarExchangeLength$ is a threshold of the $x$ (resp.~$y$) dimension over which a $y$-TW (resp. $x$-TW) transforms into a $y$-VW (resp: $x$-TW). These lines $x=7\DipolarExchangeLength$ and $t=7\DipolarExchangeLength$ are omitted for clarity on \figref{fig-diagram-simuls}.

\begin{figure}
  \begin{center}
  \includegraphics[width=133.25mm]{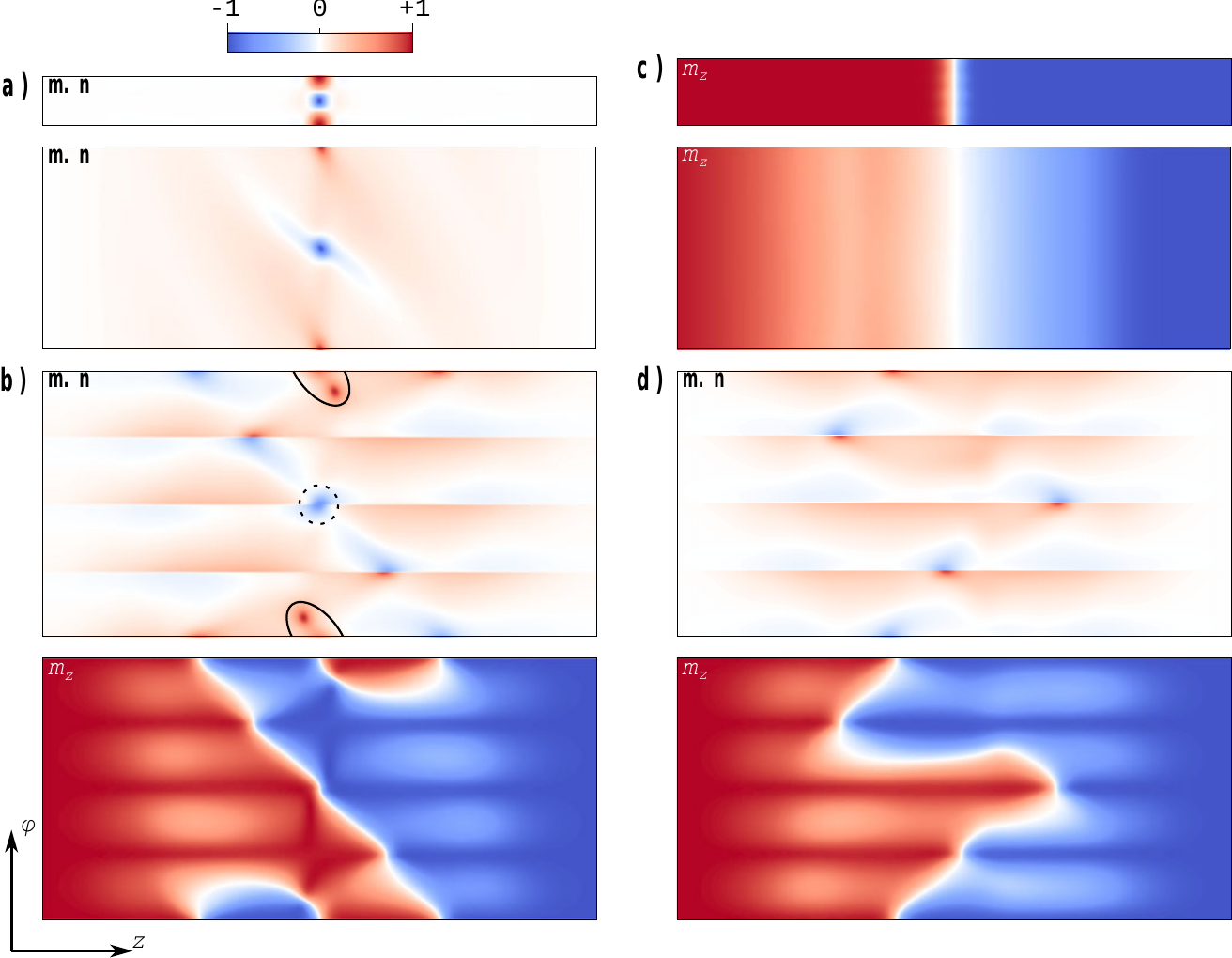}%
  \caption{\label{fig-surface-maps}\textbf{Unrolled surface maps}. In all cases the length of the view is $\unit[1000]{\nano\meter}$, and the ratio of lengths along $z$ and $\varphi$ is exact (a)~$\vect{m.n}$ maps of TVWs for disk cross-sections with diameters $\unit[30]{\nano\meter}$ and $\unit[120]{\nano\meter}$ (b)~$\vect{m.n}$ and $m_z$ maps of a TVW for square cross-section of side~$\unit[120]{\nano\meter}$. The circled (dotted-circled) areas indicate the locus of the outgoing (ingoing) flux of magnetization from the core of the TVW. We show here the sub-variety where the flux enters and exits through edges, as in \subfigref{fig-fine-points}f (c)~$m_z$ maps of a BPW for square and disk cross-sections of side~$\unit[30]{\nano\meter}$ and diameter~$\unit[120]{\nano\meter}$, respectively (d)~$\vect{m.n}$ and $m_z$ maps of a BPW for square cross-section of side~$\unit[120]{\nano\meter}$.}
  \end{center}
\end{figure}

The rise of chirality can be understood as a means to lower magnetostatic energy. The transverse curling around the core of an initially TW is similar to curling in a near-single-domain particle. The driving force for transverse curling is to reduce the magnetic charges on the sides of the wire, associated with the dipolar character of the core of the TW. Transverse curling may also be viewed as putting together two ATWs of opposite asymmetry on either side of the transverse component. This is clear when considering maps of magnetization at the surface of square wires\bracketsubfigref{fig-surface-maps}a. Absence of mirror symmetry is clear on these maps, although it is expressed through curling rather than through the entry and outlet of the flux of magnetization. By this process volume and surface charges are driven more apart one from another, further decreasing magnetostatic energy. This is why, whenever possible, transverse curling takes over asymmetry of entry/outlet points of the magnetization. As regards longitudinal curling $(\Curl\vect{m})\dotproduct{\vect u}_z$, while it is a well-recognized feature of the BPW, its relevance is less obvious for TWs at first sight, although is has been reported and sometimes called a \textsl{helical domain wall}\cite{bib-CH2012}. In fact, the physics is the same for both types of DWs: the longitudinal curling allows a progressive longitudinal variation of $m_z$ from one domain to another, in this manner spreading the volume magnetic charges $-\linepartial{m_z}{z}$ and thus decreasing magnetostatic energy. Longitudinal curling is accompanied by an increase of the DW width~$L$, as found in the simulations and well reproduced by a simple scaling law~$L\sim d^2$, see calculations in \secref{sec-scaling-laws} and \subfigref{fig-energy-wires}{b}. On this figure we computed the DW width as $L=\int_{-\infty}^{+\infty}\sin^2\theta\,\diff{z}$ following Jakubovics definition apart from a factor~2\cite{bib-JAK1978}. Notice that in disk wires with large diameter TVWs and BPWs have a very similar width, as well as similar $z$-resolved longitudinal curling\bracketsubfigref{fig-curling}{a, f-g}, confirming the common ingredient of monopolar magnetostatics and curling, independent from the DW internal structure. The variety of directions of circulation motivates our use of the word curling, which had been introduced in this context\cite{bib-FRE1957}, and is less ambiguous than the word vortex, connected to the existence of a core. We finally stress that what we consider here is the geometrical width under static conditions. While the width is often invoked as a proportionality factor determining DW mobility under applied field, it is a dynamic width that shall be considered in that case, which often differs from the static width discussed, and depends very much on the internal structure of the wall.

Finally, fine points about curling and symmetry are the following. First, in TVWs longitudinal and transverse curling compete with each other and lead to frustration, since the direction of magnetization on either side of the wire needs to be opposite for longitudinal curling, and be the same for transverse curling. This is the reason for the distorted feature of the TVW upon the rise of curling~(although the entry/exit fluxes remain symmetric)\bracketsubfigref{fig-curling}c. In practice transverse curling, characterized by antisymmetry with respect to a plane perpendicular to the wire axis, combined with the natural dihedron shape of a TW, imposes that longitudinal curling is of opposite sign on either sign of the core of the TVW\bracketsubfigref{fig-curling}{a,d-f}. This frustration is accommodated different ways depending on the wire size\bracketsubfigref{fig-curling}{d-e}, or whether of square or disk cross-section\bracketsubfigref{fig-curling}{e-f}. In the first case transverse curling is maintained and longitudinal curling yields, while it is the opposite for disk-based sections. This remark is connected to the fact that curling in square cross-sections, especially of longitudinal type, induces an extra cost of energy due to the sharp edges. This decreases the efficiency of curling to lower magnetostatic energy, and thus makes both types of DWs more costly at large diameter where the DW length increases like $d^2$. This explains why the wall width is smaller for square cross-sections compared to disks, for a given area of cross-section\bracketsubfigref{fig-energy-wires}b. It particularly increases the energy of the BPW where longitudinal curling is the strongest, making it barely favorable against the TVW in square wires\bracketsubfigref{fig-energy-wires}d. This induces the occurrence of an asymmetric BPW (ABPW), which does not exist in wires with a disk cross-section\bracketsubfigref{fig-surface-maps}d. Again, the surface maps have the same shape as ATW in flat strips. For the same reason of the cost associated with curling, ATWs may be formed in wires with square cross-section, while they do not with those of disk cross-section~(not shown here). These ATWs then prevail over symmetric TVW until above $\unit[45]{\nano\meter}$. Yet another fine feature related to magnetostatics is the rise of a slight backward longitudinal component around a BPW for disk cross-section at large diameter\bracketsubfigref{fig-surface-maps}{c}. These are the consequence of the distribution of head-to-head magnetic charge centered on the wall, and its principle is similar to concertina features is soft planar magnetic elements. These head-to-head charges are also responsible for the outward radial tilt of magnetization in the longitudinal curling\cite{bib-THI2006} of both TVWs and BPW~(see $\vect{m}\dotproduct\vect{n}$ maps on \subfigref{fig-surface-maps}{b,d}, showing an imbalance towards positive values).

\begin{figure}
  \begin{center}
  \includegraphics[width=121.211mm]{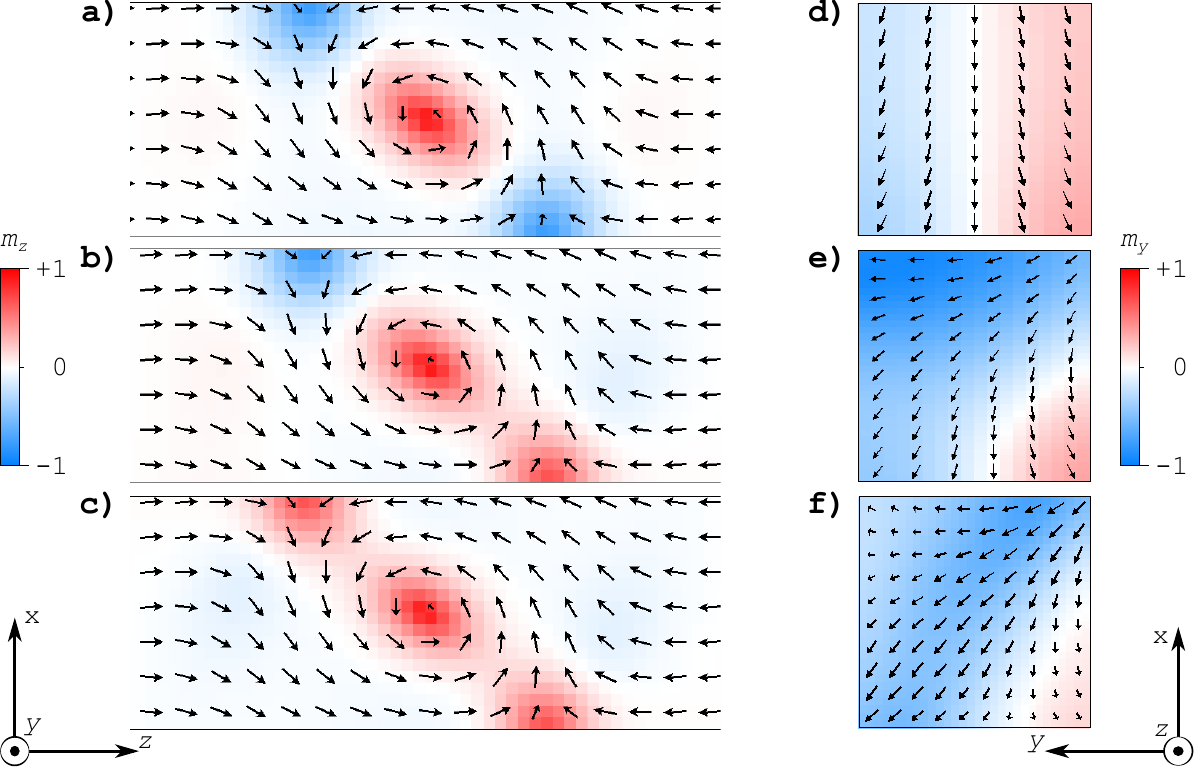}%
  \caption{\label{fig-fine-points}\textbf{Finer points} (a-c)~Three sub-varieties of VW in strips with cross-section $\unit[80\times40]{\nano\meter}$ (d-f)~Three sub-varieties of TVW in wires with square cross-section of side $\unit[30]{\nano\meter}$, $\unit[36]{\nano\meter}$ and $\unit[60]{\nano\meter}$ from top to bottom.}
  \end{center}
\end{figure}

\subsection{Fine micromagnetic features}

The above description covers the main features of domain walls in one-dimensional systems. Nevertheless finer features may occur, mostly due to the existence of flat surfaces and sharp edges. These features occur predominantly for compact cross-sections with $t\approx w$. The resulting increased number of possible states makes the phase diagram more complex to describe close to the diagonal, which explains that we deliberately discontinued some lines in the vicinity of the diagonal\bracketfigref{fig-diagram-simuls}. Besides, the transition from one type to another type of subvariety is associated with minute changes of energy for difference widths, which nonetheless are  responsible for the somewhat irregular shape of some curves, \eg the BPW/TW one.

The finer features may be of two types, shortly described below.

In strips of significant thickness the cost of magnetization perpendicular to an edge becomes prohibitive. Thus, for a VW magnetization tends to be perpendicular to the strip at the two such locus at the strip sides, for strip width larger than a few times~$\DipolarExchangeLength$. These locus are therefore sometimes described as half antivortices\cite{bib-TCH2005,bib-BRA2012}, owing to the distribution of in-plane magnetization, and the existence of a vertically-magnetized core. The question arising, is whether the core of the half antivortices shall be parallel or antiparallel to the vortex core. Various varieties may be found as (meta)stable states for the same geometry, and reached depending on the initial computing conditions: antiparallel, parallel or even one anti-vortex in each direction\bracketsubfigref{fig-fine-points}{a-c}. For large and thin strips, \ie when the DW has a predominant VW feature, the state with lowest energy tend to be with anti-vortices antiparallel to the core of the vortex. The driving force for this is minimization of magnetostatic coupling, while exchange is negligible. For narrower or thicker strips, \ie when the TW feature of the DW is growing, the state of lower energy tends to be with anti-vortices parallel to the core of the domain wall. The driving force is that by gaining features of a TW, the core grows in size and tends to form a continuous sheet of transverse magnetization, in which switching the direction of magnetization would imply a high cost in exchange energy.

At sharp edges of a 3D magnetic element the density of magnetostatic energy diverges logarithmically\cite{bib-RAV1998}. Deviations from uniform magnetization lift the degeneracy expected from demagnetization coefficients calculated for a uniformly-magnetized system. Magnetization prefers to point along the bisector or perpendicular to it, both possibilities being separated by an energy barrier. This has been recognized in near single domain bulk\cite{bib-RAV1998b} and thin film elements\cite{bib-COW1998c} to give rise to so-called leaf or flower states. The same physics may occur, applied to the transverse core of a TVW. The flux of magnetization may cross the strip from one side to the opposite side, or from one edge to the opposite edge, or as an intermediate situation from one side to one of the opposite edges\bracketsubfigref{fig-fine-points}{d-f}. The trend seems that the state of lowest energy if the side-to-side configuration for smaller lateral dimensions, and the edge-to-edge configuration for larger lateral dimensions. It has not been attempted to provide a full picture of the phenomenon, due to the large number of states involved. Indeed, this issues holds for both VW and TW, giving rise to numerous sub-varieties.

\section{Analytical scaling laws}
\label{sec-scaling-laws}

In this section we derive scaling laws pertaining to a few aspects of the phase diagram. The scaling laws are not intended to be rigorous nor numerically accurate, however to provide trends and physical insights to the micromagnetic simulations. In the following we will write $\Kd=\muZero\Ms^2/2$ the dipolar constant.

\subsection{Transverse versus vortex walls in flat strips}

Here we consider the TW/VW iso-energy line in a flat strip already reported, for which a phenomenological law was mentioned in previous works\cite{bib-MIC1997,bib-NAK2005}. First, we notice on the simulation results that the length of the $\pi/2$ sub-walls is identical in both cases. Their energy can therefore be disregarded for the difference. As regards magnetostatic energy, MFM shows that the total head-to-head charge $Q=2\Ms tw$ is spread over an area $S_\mathrm{VW}=2w^2$ for the VW and $S_\mathrm{TW}=w^2$ for the TW\cite{bib-CHA2010}. The areal density of charge $\sigma=Q/S$ amounts to  $\sigma_\mathrm{VW}=\Ms t/w$ and $\sigma_\mathrm{TW}=2\Ms t/w$. The resulting magnetostatic energy can be estimated as $(1/2)\muZero\Hd^2$ integrated over area $S$ and height $w$ (used as a cut-off) on either side of the strip, with $\Hd\approx\sigma/2$. This yields $\mathcal{E}_\mathrm{d,VW}=\Kd t^2w$ and $\mathcal{E}_\mathrm{d,TW}=2\Kd t^2w$. This shows that the TW has a higher magnetostatic energy of head-to-head origin, however the VW also holds extra exchange and magnetostatic energy, related to the vortex core and its surroundings. As the core has a diameter around $3\DipolarExchangeLength$\cite{bib-FEL1965b}, its energy itself is of the order of $2\pi(3\DipolarExchangeLength/2)^2 \Kd t$, taking into account an equipartition of energy of exchange with dipolar with the front factor~2. Finally, exchange energy integrated from around the core to the strip edge may be estimated at $2\pi t A \ln(2w/3\DipolarExchangeLength)$ based on the cut-off radius $3\DipolarExchangeLength/2$. Putting everything together and substituting the numerical value 2 to 3 for the logarithm as valid for the range of geometries studied in \cite{bib-NAK2005}, one finds the following equation for the iso-energy line:

\begin{equation}\label{eqn-scaling-tw-vw}
  tw\sim\DipolarExchangeLength^2
\end{equation}

\noindent This provides the good scaling law proposed in the previous reports, with a surprisingly good numerical value~($\approx30$) compared to the one fitted to the simulation data~($\approx61$).

\subsection{Transverse versus asymmetric transverse walls in flat strips}

Examination of TWs and ATWs in strips shows that their difference is essentially restricted close to the edge of the strip, on the large side of the mostly triangular transverse component~(upper side on \subfigref{fig-tw-vw-strip}{d-e}). We therefore focus on this area to evaluate the difference of energy between the two walls. In the TW magnetization is perpendicular to the edge over an edge length of the order of $w$. Considering the edge as a plane with a charge density $\Ms$, implies a magnetostatic cost $\Delta\mathcal{E}_\mathrm{d}\sim\Kd wt^2$, based on arguments similar to those derived in the previous paragraphs. The cost associated with the ATW comes from several aspects. The progressive rotation of magnetization from the center of the strip to the edge implies a cost of density of exchange energy $\sim A/w^2$. Considered over part of the width~$w$, over thickness~$t$ and edge length $\sim w$ implies a total cost $\sim At$. Another cost is associated with the pinch of magnetization flux at one corner of the triangle. The associated exchange energy scales with $At$, while the associated magnetostatic energy may be estimated from that of a sphere with an internal charge $Q\approx\pi\DipolarExchangeLength t\Ms$, $\pi\DipolarExchangeLength$ being used as a cutoff for the concentration of flux. This localized charge becomes increasingly clear for larger thickness\bracketsubfigref{fig-landau-wall}{a,b}, and is also evidenced with MFM\cite{bib-CHA2010}. The magnetostatics of the charged sphere implies a cost $\sim t^2\DipolarExchangeLength\Kd$, with a radius set to~$\DipolarExchangeLength$. Using this cutoff rather that thickness is required to fit the cutoff $\DipolarExchangeLength$ introduced to estimate the charge within the sphere. In the end, $\Delta\mathcal{E}_\mathrm{ATW}\sim At+ t^2\DipolarExchangeLength\Kd$, omitting numerical factors. Equating $\Delta\mathcal{E}_\mathrm{TW}$ and $\Delta\mathcal{E}_\mathrm{ATW}$ provides the following scaling law:

\begin{equation}%
  \label{eqn-scaling-tw-atw}%
  (w-w_0)t\sim \DipolarExchangeLength^2
\end{equation}

\noindent with $w_0\sim\DipolarExchangeLength$. For large width this law agrees well with the known phase diagram\bracketsubfigref{eqn-scaling-tw-vw}. The offset $w_0$ explains why the ATW/TW curve, although well below the TW/VW one for large diameter, seemed to cross it towards smaller diameters. Our numerical results confirm this crossing and the then rapid increase in thickness\bracketfigref{fig-diagram-simuls}. It is also consistent with the fact that this line is not expected to reach the $y$ axis, based on topology and symmetry arguments\bracketsubfigref{fig-diagrams}{b}.

\subsection{Length and energy of domain walls in cylindrical wires}

We consider a domain wall in wires with a disk cross-section, and are interested in its energy and geometrical width at rest. We search for scaling laws and not numerical coefficients, so that the exact definition of DW width is not important (Lilley\cite{bib-LIL1950}, Jakubovics\cite{bib-JAK1978}, Thiele\cite{bib-THI1973} or other). It was noticed by Nakatani \etal that for very small radius ($R\ll\sqrt{A/\Kd}$) the 1d model is a good approximation for the DW energy and width, using as anisotropy energy $\Kd/2$, related to the $N=1/2$ demagnetizing coefficient across the wire\cite{bib-NAK2005}. A refined model was proposed later to derive the diameter dependence of this law, based on a variational model\cite{bib-HER2014}. However this model is valid only for small diameters as it disregards the monopolar charge of the DW. Here we propose a scaling law for large radius. The total head-to-head charge in the DW is $Q=2\pi R^2 \Ms$. Assuming that this charge is uniformly distributed in a spherical volume of radius $R$ the associated total magnetostatic energy $\mathcal{E}_\mathrm{d}$ is expected to scale like $\muZero Q^2/R$, so: $\mathcal{E}_\mathrm{d}\sim\Kd R^3$. As the integrated exchange energy of a non-uniform state in such a volume scales like $R$\cite{bib-FEL1965,bib-THI2003}, this shows that very rapidly magnetostatic energy becomes the dominant term upon increasing $R$. To decrease $\mathcal{E}_\mathrm{d}$, the width of the wall (\ie, along the wire axis) will tend to increase. In practice a balance will be found with exchange energy associated with this stretching, which we evaluate below.

We consider that $Q$ is spread over a length $L$, so that the volume density of charges is $\rho=Q/\pi R^2 L$. Leaving aside numerical factors, and computing magnetostatics as for an infinitely-long cylinder, one finds $\mathcal{E}_\mathrm{d}\sim \Kd R^4/L$ (note the consistency with the above case when $L$ is set to~$R$). To reach this the logarithmic divergence is bound with cutoff at a radius $L$ from the axis. We have seen in the simulations that magnetization tends to curl around the axis to avoid the formation of surface charges while progressively changing the component of magnetization along the axis of the wire. This implies a total cost of exchange energy $\mathcal{E}_\mathrm{ex}$ scaling like $LR^2 (A/R^2)$, thus: $\mathcal{E}_\mathrm{ex}\sim AL$. Minimizing $\mathcal{E}=\mathcal{E}_\mathrm{d}+\mathcal{E}_\mathrm{ex}$ yields

\begin{eqnarray}
  \label{eqn-scaling-large-width}%
  \mathcal{E} &\sim& A R^2/\DipolarExchangeLength \\
  \label{eqn-scaling-large-energy}%
  L           &\sim& R^2/\DipolarExchangeLength
\end{eqnarray}

The wall length is therefore expected to increase rapidly with the wire radius, confirming existing simulations\cite{bib-ALL2009}. Quantitatively, this scaling law fits well our own simulations\bracketsubfigref{fig-energy-wires}b. The cross-over from the low-diameter regime to the $R^2$ law may be defined as the intercept of the asymptote of the latter, with the $x$~axis. This occurs around $\unit[40]{\nano\meter}\approx8\DipolarExchangeLength$. This happens to be close to the onset of curling, confirming it as the key ingredient. The $R^2$ scaling law for energy is also in fair agreement with simulations\bracketsubfigref{fig-energy-wires}a.

\section{Conclusion and trends}

Through micromagnetic simulations and analytical scaling laws we considered head-to-head magnetic domain walls in one dimensional structures, with geometry spanning from thin strips to wires with a square or disk cross-section. The former are experimentally relevant for structures made by lithography, while the latter are more relevant for bottom-up synthesis. All domain walls found fall into only two varieties, based on their topology: transverse vortex walls~(TVWs) and Bloch-point walls~(BPWs). For wires the former display both a transverse (flux of magnetization going through the wire) and a vortex (also named transverse curling) feature. When the geometry goes towards thin strips the transverse or vortex feature takes over and yield the already known transverse and vortex walls, depending on the direction of the through flux, in-plane or out-of-plane, respectively. Concerning BPWs, it is found that they exist for non-perfectly-disk- or -square-shaped wires, and may even be the ground state over a significant range of geometries. These are relevant for thick strips made by top-down techniques, so the BPW and its predicted peculiarities of magnetization dynamics may be of more general relevance than previously thought. We deliver a phase diagram of the different types of domain walls, based on their energy and discussed in terms of phase transitions of either first or second order. The latter are concerned with textures of magnetization developed to close as much as possible the flux, and spread the head-to-head charges along the axis of the structure, to lower magnetostatic energy. These textures develop when at least one of the transverse dimensions equals seven times the dipolar exchange length $\sqrt{2A/\muZero\Ms^2}$. They may take the form of either an asymmetry as in the already known case of asymmetric transverse walls in flat strips, or curling (with both transverse and longitudinal components). Curling generally yields a more efficient decrease of energy than asymmetric domain walls, except for some cases of wires with square cross-section, because curling is associated with magnetostatic energy at edges.

The theoretical and experimental consideration of magnetic domain walls in thin strips is now rather extensive, following about twenty years of reports. To the contrary, domain walls in wires (\ie with compact cross-section such as square or disk) have mostly been described theoretically, either through analytics or simulations. Since pioneering works around 2000 several tens of papers have been devoted to the expected types, energetics and peculiar dynamics of motion under magnetic field or spin-polarized current. The first experimental results on the statics of such domain walls are emerging\cite{bib-BIZ2013,bib-FER2013,bib-FRU2014}. Investigating their control in wires and tubes and their dynamics is a challenging however timely direction of research. These hold both exciting prospects for fundamental findings such as velocities up to $\unit[1]{\kilo\meter\per\second}$\cite{bib-THI2006} or interaction with spin waves\cite{bib-YAN2011b}, and are directly applicable to the proposal for a three-dimensional race-track memory making use of densely-packed arrays of magnetic wires\cite{bib-PAR2008}. Our extended phase diagram shows that the physics of domain walls in wires, especially associated with the exotic Bloch-point domain wall, may also be searched for in thick strips, and thus be achievable with top-down techniques.

\section{Further information}

Most of the above results are previously unpublished. Nevertheless, below are a few key references and reviews concerning pre-existing knowledge. R.~McMichael and M.~Donahue\cite{bib-MIC1997}, and Y.~Nakatani \etal\cite{bib-NAK2005} provided the theoretical overview of domain walls in flat strips. The theoretical investigation of domain walls in wires was pioneered by R.~Hertel\cite{bib-HER2002a,bib-NIE2002} and H.~Forster\cite{bib-FOR2002,bib-FOR2002b}. Excellent reviews (both literature and original research) for their statics and dynamics under magnetic field and spin-polarized current can be found in two book chapters by A.~Thiaville and Y.~Nakatani\cite{bib-THI2006,bib-THI2008}. Many of the concepts developed also apply to the geometry of tubes, published later by several authors.

\section*{Acknowledgements}

The research leading to these results has received funding from the European Unions's 7th Framework Programme under grant agreement n°309589 (M3d). We acknowledge helpful discussions with Ch.~Thirion and J.~Vogel (Institut NEEL-Grenoble), A.~Fernandez-Pacheco (Cavendish Laboratory-Cambridge), and critical reading of the manuscript by A.~Thiaville (LPS-Orsay) and R.~Sch\"{a}fer (IFW-Dresden).


\end{document}